\documentclass[twocolumn]{aastex6}

\usepackage{siriodef}
\usepackage{capt-of}
\usepackage{graphicx}
\usepackage{subfigure}

\shortauthors{Belli et al.}
\submitted{Accepted for publication in ApJ Letters}

\begin{document}

\title{\kmostd\ reveals low-level star formation activity \\ in massive quiescent galaxies at $0.7 < \MakeLowercase{z} < 2.7$}
\thanks{Based on observations collected at the European Southern Observatory under programs 092.A-0091, 093.A-0079, 094.A-0217, 095.A-0047, 096.A-0025, and 097.A-0028}

\author{Sirio Belli\altaffilmark{1},
Reinhard Genzel\altaffilmark{1,2,3},
Natascha M. F\"orster Schreiber\altaffilmark{1},
Emily Wisnioski\altaffilmark{1},
David J. Wilman\altaffilmark{1,5},
Stijn Wuyts\altaffilmark{4},
J. Trevor Mendel\altaffilmark{1},
Alessandra Beifiori\altaffilmark{1,5},
Ralf Bender\altaffilmark{1,5},
Gabriel B. Brammer\altaffilmark{6},
Andreas Burkert\altaffilmark{1,5},
Jeffrey Chan\altaffilmark{1,5,7},
Rebecca L. Davies\altaffilmark{1},
Ric Davies\altaffilmark{1},
Maximilian Fabricius\altaffilmark{1,5},
Matteo Fossati\altaffilmark{1},
Audrey Galametz\altaffilmark{1},
Philipp Lang\altaffilmark{1},
Dieter Lutz\altaffilmark{1},
Ivelina G. Momcheva\altaffilmark{6},
Erica J. Nelson\altaffilmark{1},
Roberto P. Saglia\altaffilmark{1,5},
Linda J. Tacconi\altaffilmark{1},
Ken-ichi Tadaki\altaffilmark{1},
Hannah \"Ubler\altaffilmark{1},
Pieter van Dokkum\altaffilmark{8},
}



\begin{abstract}
We explore the \Halpha\ emission in the massive quiescent galaxies observed by the \kmostd\ survey at $0.7 < z < 2.7$. The \Halpha\ line is robustly detected in 20 out of 120 $UVJ$-selected quiescent galaxies, and we classify the emission mechanism using the \Halpha\ line width and the \NII/\Halpha\ line ratio. We find that AGN are likely to be responsible for the line emission in more than half of the cases. We also find robust evidence for star formation activity in nine quiescent galaxies, which we explore in detail. The \Halpha\ kinematics reveal rotating disks in five of the nine galaxies. The dust-corrected \Halpha\ star formation rates are low ($0.2 - 7$ \Msun/yr), and place these systems significantly below the main sequence. The 24$\mu$m-based infrared luminosities, instead, overestimate the star formation rates. These galaxies present a lower gas-phase metallicity compared to star-forming objects with similar stellar mass, and many of them have close companions. We therefore conclude that the low-level star formation activity in these nine quiescent galaxies is likely to be fueled by inflowing gas or minor mergers, and could be a sign of rejuvenation events.
\end{abstract}

\keywords{galaxies: high-redshift --- galaxies: evolution --- galaxies: ISM --- galaxies: star formation}


\section{Introduction}
\label{sec:introduction}

\footnotetext[1]{Max-Planck-Institut f\"ur Extraterrestrische Physik (MPE), Giessenbachstr. 1, D-85748 Garching, Germany \vspace{-1mm}}
\footnotetext[2]{Department of Physics, Le Conte Hall, University of California, Berkeley, CA 94720, USA \vspace{-1mm}}
\footnotetext[3]{Department of Astronomy, Campbell Hall, University of California, Berkeley, CA 94720, USA \vspace{-1mm}}
\footnotetext[4]{Department of Physics, University of Bath, Claverton Down, Bath, BA2 7AY, UK \vspace{-1mm}}
\footnotetext[5]{Universit\"ats-Sternwarte, Ludwig-Maximilians-Universit\"at, Scheinerstrasse 1, D-81679 M\"unchen, Germany \vspace{-1mm}}
\footnotetext[6]{Space Telescope Science Institute, 3700 San Martin Drive, Baltimore, MD 21218, USA \vspace{-1mm}}
\footnotetext[7]{Department of Physics and Astronomy, University of California, Riverside, CA 92521, USA \vspace{-1mm}}
\footnotetext[8]{Astronomy Department, Yale University, New Haven, CT 06511, USA \vspace{-1mm}}

The discovery of a population of massive quiescent galaxies at high redshift \citep[e.g.,][]{franx03, cimatti04, labbe05, kriek06} has important implications for the early phases of galaxy evolution. The formation of these objects requires a rapid assembly of stellar mass, after which the star formation activity terminates suddenly. The nature of such \emph{quenching} of star formation, which is one of the most important drivers of galaxy evolution, is largely unknown.

Furthermore, it is unclear how star formation remains quenched throughout the lifetime of a quiescent galaxy, in spite of the continuous replenishment of gas due to inflows and mass loss from stellar evolution. Theoretical studies have suggested several physical processes that may be responsible for heating the gas and preventing further star formation, such as low-level feedback from active galactic nuclei (AGN) \citep[e.g.,][]{croton06}, stellar winds from old stars \citep[e.g.,][]{conroy15}, or gravitational heating \citep[e.g.,][]{dekel08}. One of the key predictions of the models is the level of residual star formation in quiescent systems, which therefore represents a fundamental goal for observational studies.

\begin{figure*}[tbp]
\centering
\includegraphics[width=0.94\textwidth]{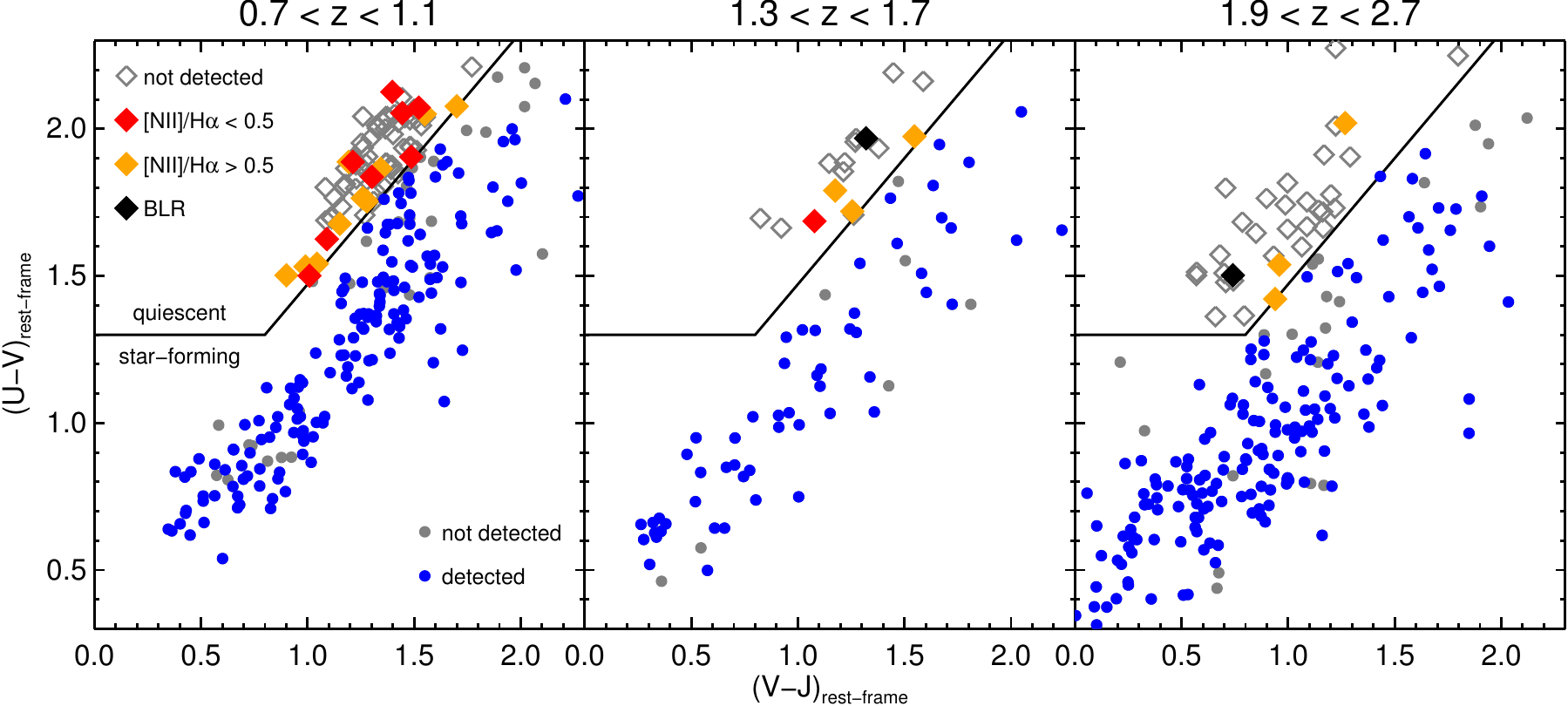}
\caption{$UVJ$ diagram for the \kmostd\ sample, split into redshift bins. The solid line separates quiescent (diamonds) from star-forming galaxies (circles). Filled diamonds mark quiescent galaxies detected in \Halpha, with colors indicating the type of emission. Star-forming galaxies detected in \Halpha\ are shown in blue.}
\label{fig:UVJ}
\end{figure*}

At high redshift, where gas accretion plays an increasingly important role, quiescent galaxies have been studied mainly via photometric data. Fitting templates to the observed spectral energy distribution (SED) generally yields a reliable measurement of the stellar mass, but less so for the star formation rate, particularly for low levels of star formation or when dust extinction is present \citep[for a review, see][]{conroy13review}. An alternative technique is the analysis of the infrared emission by dust. However, this method is hampered by the unknown contribution of different sources to the heating of dust in quiescent galaxies, and can only be performed via stacking due to the low sensitivity of infrared observations \citep{fumagalli14, man16}.

The most direct way to measure star formation activity is by observing the \Halpha\ emission line. In this Letter we analyze the \Halpha\ and \NII\ emission in 120 quiescent galaxies at $0.7 < z < 2.7$ using spatially resolved spectroscopic data from the \kmostd\ survey. Due to the large sample and the deep exposures, this dataset allows us for the first time to study in detail the emission line properties for a substantial number of individual quiescent galaxies at high redshift.

Throughout the paper we assume a \citet{chabrier03} initial mass function (IMF) and a $\Lambda$CDM cosmology with $\Omega_M$=0.3, $\Omega_{\Lambda}$=0.7 and $H_0$= 70 km s$^{-1}$ Mpc$^{-1}$.


\section{Data}
\label{sec:data}

The present analysis is based on data from \kmostd, an ongoing GTO program with the KMOS spectrograph \citep{sharples13} at the Very Large Telescope. The details of the survey, including design, observations, and data reduction, are described in \citet{wisnioski15}; here we only give a brief summary.

\kmostd\ is a large survey aimed at collecting resolved spectroscopic observations for over 600 high-redshift galaxies, using the $2\farcs8 \times 2\farcs8$ near-infrared integral field units of KMOS. The observations are carried out in three of the CANDELS fields \citep{grogin11, koekemoer11}, for which a wealth of ancillary data are publicly available. Stellar masses, dust extinction, and rest-frame colors are derived following a standard SED modelling, for which the details are presented in \citet{wuyts11}.

The spectroscopic targets are selected in three redshift bins, $0.7 < z < 1.1$, $1.3 < z < 1.7$, and $1.9 < z < 2.7$, using spectroscopic redshifts from the literature (available for about half of the sample) or the grism-based redshifts from the 3D-HST catalog \citep{brammer12, momcheva16}. We apply a magnitude cut $K_S < 23 $, which yields a 95\% mass-complete sample above $\log \Mstar \sim $ 9.7, 10.2, and 10.5 in the three bins, resulting in a diverse target sample that spans a wide range in masses, environments, sizes, colors, and star formation rates.

The observations were obtained in natural seeing, with a spatial resolution of $0\farcs4 - 0\farcs8$ FWHM (median $0\farcs53$) and a spectral resolution $\sigma \sim 25-50$ km~s$^{-1}$. Exposure times of individual objects vary between 4 and 25 hours. In this work we include the 560 galaxies observed in the first 3 years of the survey, up to September 2016.

For each target we extract a 1D spectrum by coadding the spaxels associated with the galaxy, and visually identify candidate emission lines. Based on these spectra, we identify 399 robust \Halpha\ detections, i.e. with a signal-to-noise ratio of at least 3 and no strong contamination from sky line residuals. For the objects where the data cube reveals spatially resolved motion, we derive a velocity map, correct the data cube for the resolved velocity gradients, and extract a velocity-corrected spectrum. In the present analysis we use such velocity-corrected spectrum for galaxies with spatially resolved \Halpha\ emission; for the remainder of the sample we adopt the simple extraction.


\section{\Halpha\ Emission in Quiescent Galaxies}
\label{sec:emlines}

We use the rest-frame $U-V$ and $V-J$ colors to select quiescent galaxies. This method is remarkably effective even when dust reddening is present \citep[e.g.,][]{wuyts07, williams09}. In Figure \ref{fig:UVJ} we show the \kmostd\ target sample on the $UVJ$ diagram, split into the three redshift bins. The black line marks the selection box for quiescent galaxies. We adopt the criteria of \citet{muzzin13}, without the $V-J < 1.5$ requirement, which would exclude very red quiescent systems \citep[e.g.,][]{vanderwel14, whitaker15}.

The galaxies with a robust detection of \Halpha\ in emission are shown with colored symbols. Star-forming galaxies are detected with an overall success rate of 86\%. As expected, quiescent galaxies have a lower detection rate. Out of 120 $UVJ$-quiescent galaxies, 20 objects are robustly detected in \Halpha, corresponding to a detection rate of 17\%, which increases to 27\% when including 13 marginal or contaminated detections. It is possible that we missed a small number of emission-line targets for which \Halpha\ is hidden by sky lines, falls outside the observed wavelength range, or is buried in a strong \Halpha\ absorption line. Most of the detections happen to lie near the edge of the $UVJ$ selection box, an important point that we will discuss further. Also, the detection rate is significantly higher in the lowest redshift bin, likely because of cosmological dimming.

\begin{figure}[tbp]
\includegraphics[width=0.43\textwidth]{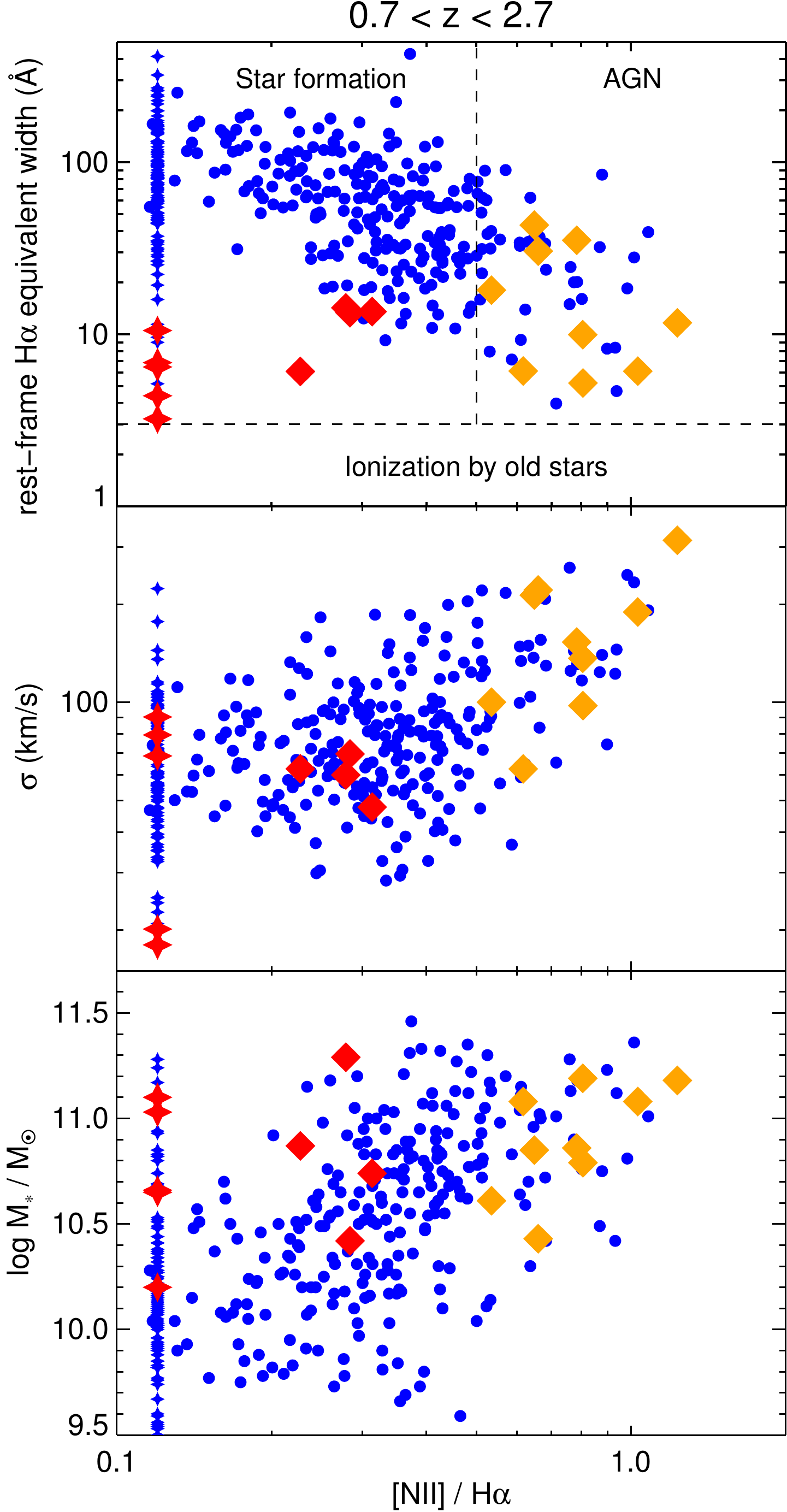}
\caption{Rest-frame \Halpha\ equivalent width (top), velocity dispersion (center), and stellar mass (bottom) as a function of the \NII/\Halpha\ flux ratio. Symbols as in Figure~\ref{fig:UVJ}; stars mark those galaxies for which [NII] is not detected, to which we assign an arbitrary value of [NII]/\Halpha$ = 0.12$.}
\label{fig:whan}
\end{figure}

The emission line properties vary significantly among the detected quiescent galaxies. We explore this quantitatively by fitting Gaussians to \Halpha\ and \NII$6583$ and measuring line widths and fluxes. In two cases (shown as black diamonds in Figure \ref{fig:UVJ}) the \Halpha\ line is very broad ($\sigma \sim 500 - 1000$ km~s$^{-1}$), and \NII\ is not present. This emission likely originates from the broad line region (BLR) around a central black hole. Both galaxies are detected in the X-rays, and we exclude them from further analysis. We split the remaining objects into two subsamples, according to the measured \NII/\Halpha\ flux ratio, which can be used to discriminate between star formation and other emission mechanisms. Photoionization models show that star formation activity produces emission lines characterized by \NII/\Halpha\ $ < 0.5$ \citep[e.g.,][]{kewley13}; the presence of shocks or AGN activity increases the line ratio. We find that nine galaxies have weak or non-detected \NII\ emission (\NII/\Halpha\ $ < 0.5$, hereafter \NII-weak galaxies). The remaining nine galaxies (hereafter \NII-strong) have \NII/\Halpha\ $ > 0.5$, and are thus formally inconsistent with an ionization field due to star formation alone. Of the additional 13 galaxies for which \Halpha\ is marginally detected or is contaminated by a strong sky lines, we classify six as [NII]-weak. The other seven are \NII-strong and show at least two lines among \NII$6548$, \Halpha, and \NII$6583$. The simultaneous detection of two emission lines with the correct wavelength ratio ensures that the redshifts of these seven objects are secure.

We investigate further the nature of the emission in Figure \ref{fig:whan}. The top panel shows the rest-frame \Halpha\ equivalent width\footnote{The equivalent width is calculated dividing the line flux by the continuum of the best-fit SED model, taking into account the underlying absorption.} as a function of the \NII/\Halpha\ line ratio for the \kmostd\ sample. This so-called WHAN diagram is useful for separating galaxies according to the emission mechanism \citep{cidfernandes11}. The figure confirms that the \NII-weak galaxies in our sample populate the star forming region. It also rules out another possible explanation for the observed emission, i.e., the ionization due to old stars, since all rest-frame equivalent widths are above 3 \AA, which has been proposed as the threshold value in the local universe \citep{cidfernandes11, belfiore16}.

\begin{figure}[tbp]
\centering
\includegraphics[width=0.45\textwidth]{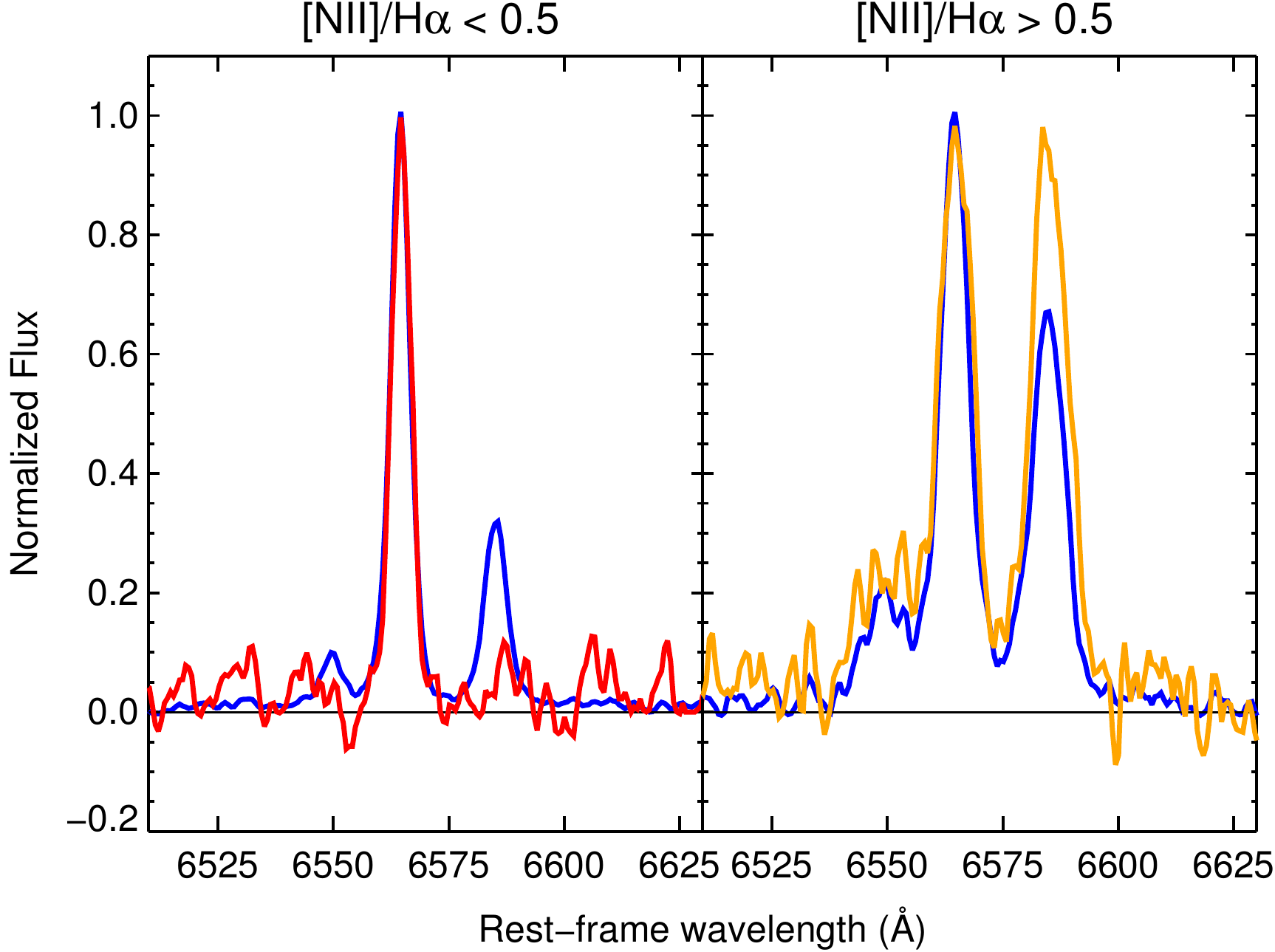}
\caption{Stacked spectra of \Halpha-detected galaxies, split into the \NII-weak (left) and \NII-strong (right) populations. In each panel, the red or orange spectrum is the stack of $UVJ$-quiescent galaxies, while the blue spectrum is the stack of the star-forming galaxies more massive than $\log \Mstar/\Msun = 10.4$ (to which we apply the same cut in \NII/\Halpha\ ratio). Before stacking, each spectrum is continuum-subtracted and flux-normalized.\vspace{-3mm}}
\label{fig:stacks}
\end{figure}

The central panel of Figure \ref{fig:whan} shows the intrinsic (i.e., corrected for instrumental resolution and, when possible, resolved motion) velocity dispersion as a function of the line ratio. At high values of \NII/\Halpha\ all galaxies, whether they are star-forming or quiescent, follow the same trend of larger velocity dispersions at higher line ratios, which suggests the presence of shocks \citep[e.g.,][]{dopita95}. Interestingly, the \NII-weak quiescent galaxies have consistently smaller velocity dispersions. Since the \NII-weak and \NII-strong populations have a similar stellar mass distribution (as is clear from the bottom panel of Figure \ref{fig:whan}), we conclude that the difference in velocity dispersion is genuine and not due to a difference in stellar mass.

Finally, we stack the spectra for the two populations, including the marginal detections, and plot them in Figure \ref{fig:stacks}, along with the stack of all the star-forming galaxies in \kmostd\ with stellar masses in the same range as the quiescent targets ($\log \Mstar/\Msun > 10.4$). The stacks show that \NII-weak quiescent and star forming galaxies have a similar \Halpha\ line width, which is smaller than that of \NII-strong galaxies.

The observed \NII/\Halpha\ ratios suggest that the line emission in the \NII-strong objects is due to AGN activity. As additional evidence for the presence of AGN, three out of nine of these galaxies are detected in the X-rays, as opposed to zero out of nine for the \NII-weak sample. Furthermore, the observed trend between velocity dispersion and line ratio indicates the presence of shocks. One possibility is that the \NII-strong emission is due to large-scale outflows driven by AGN activity, a scenario that is also consistent with the fact that we spatially resolve some of the emission lines in the KMOS data cubes. This type of outflow emission has been already observed by \citet{forsterschreiber14} in the form of broad nuclear emission in star-forming galaxies, albeit with higher line widths. We note the presence of a similar broad component in the stack of the \NII-strong sample, shown in the right panel of Figure \ref{fig:stacks}. The larger sample of galaxies with broad emission presented by \citet{genzel14outflow} includes four galaxies with negligible star formation rates, demonstrating that nuclear outflows can also be present in quiescent galaxies. Further analysis of the \NII-strong objects will be included in a future study of nuclear outflows.


\section{Star Formation in Quiescent Galaxies}
\label{sec:starformation}

The nine \NII-weak quiescent galaxies in which we detected star formation activity offer a unique perspective on galaxy quenching near the peak of cosmic star formation history. In this section we use imaging, photometric, and spectroscopic data to explore the physical properties of these objects.

\subsection{Rest-frame Colors}

It is striking that all the quiescent galaxies hosting star formation activity except one (COS4-03894) lie around the edges of the red sequence in the $UVJ$ diagram (see Figure \ref{fig:UVJ}). In particular, most of them are found in one of the two extremities of the red sequence: three galaxies have very blue colors ($U-V \sim 1.6$), which are associated with young ages of about 1 Gyr, as confirmed by spectroscopic studies \citep{whitaker13, belli15, mendel15}; and three galaxies have very red colors ($U-V \sim 2.1$), which are likely due to a combination of old stellar ages and dust reddening.

\subsection{Morphology and Environment}

The \HST\ images of the nine galaxies, shown in Figure \ref{fig:sample}, are surprisingly rich in morphological features, with many objects having one or more close companions. In the data cubes of COS4-21030, COS4-00970, and GS3-24837 we detect \Halpha\ emission from both the main target and a satellite within 800 km~s$^{-1}$, thus confirming that the galaxies are indeed interacting with their companions.

Possible satellites beyond the limited size of the KMOS IFU are identified from the 3D-HST catalog within 50 kpc and 1-sigma redshift errors. In total seven out of the nine \NII-weak quiescent galaxies have possible lower-mass companions. We also explored whether the \Halpha\ detection correlates with the local overdensity \citep[measured by][]{fossati17}, but did not find any significant trend.

\subsection{Kinematics}

\begin{figure}[htbp]
\centering
\includegraphics[width=0.44\textwidth]{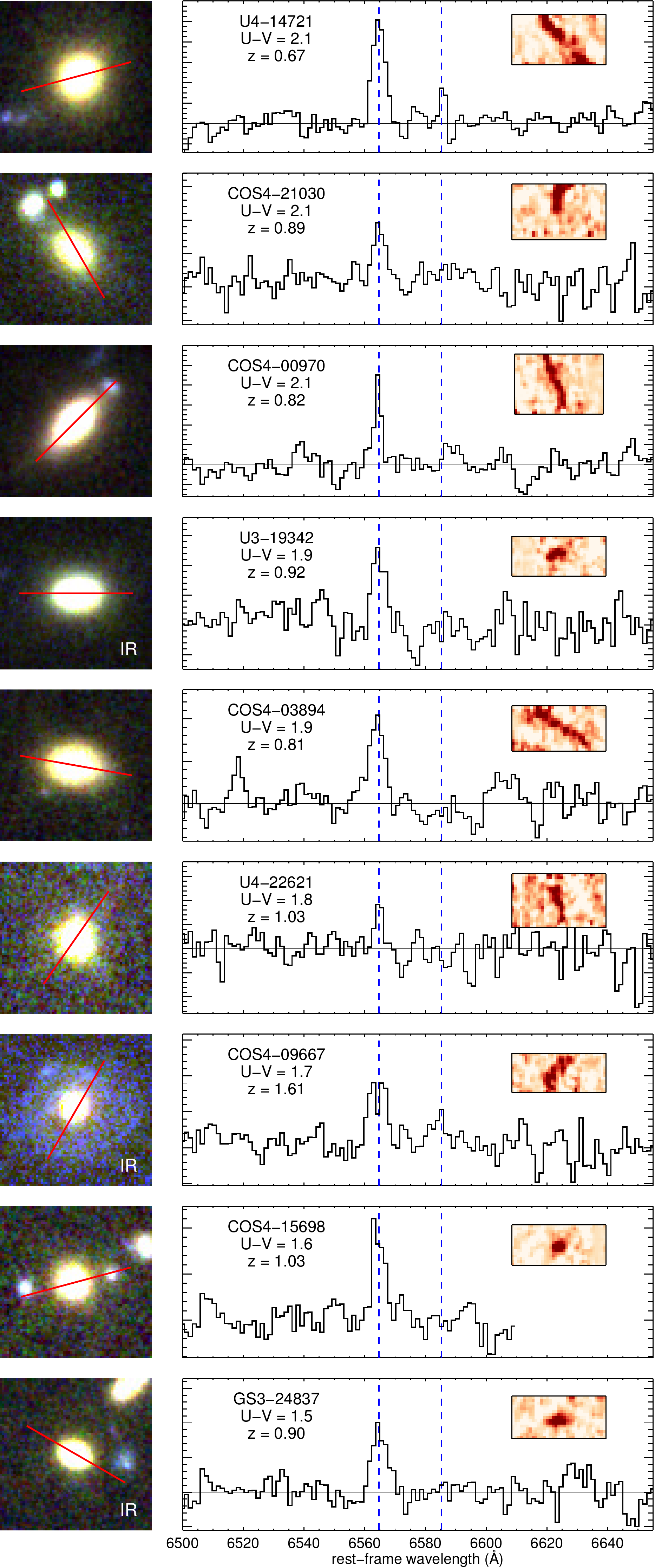}
\caption{The sample of nine \NII-weak quiescent galaxies, for which the \Halpha\ emission is due to star formation activity. Left: composite $4''\times4''$ $IJH$ image stamp from \HST. The pseudoslit, which for the spatially resolved objects is aligned with the kinematic axis, is shown in red; galaxies detected at 24$\mu$m are marked ``IR''. Right: 1D spectrum corrected for resolved motion, after suppression of sky lines and subtraction of the continuum. Blue dashed lines mark the position of \Halpha\ and \NII. For each object we list the ID, rest-frame $U-V$ color, and spectroscopic redshift. The inset shows the 2D spectrum, centered on \Halpha, extracted along the pseudoslit and continuum-subtracted.}
\label{fig:sample}
\end{figure}

In Figure \ref{fig:sample} we show the 2D spectra extracted from pseudo-slits aligned with the kinematic axes. In five of the nine galaxies the \Halpha\ emission is spatially resolved and clearly exhibits a smooth velocity gradient consistent with the presence of a gas disk. Since the median spatial resolution of our observations is $0\farcs53$, the extent of the disks must be at least $\sim4$ kpc. The rotational velocities are in the range $100 - 200$ km~s$^{-1}$, with the notable exception of COS4-03894, which presents a velocity of $\sim 400$ km~s$^{-1}$. Interestingly, this is also the only \NII-weak galaxy that in the $UVJ$ diagram lies on the central part of the red sequence, instead of being near the edge.

The integrated \Halpha\ velocity dispersions (not corrected for resolved rotation) range from 50 to 270 km~s$^{-1}$, with a median value of 103 km~s$^{-1}$. These line widths are significantly smaller than the integrated stellar velocity dispersions usually measured at these redshifts and stellar masses \citep[$\sigma \sim 150 - 300$ km~s$^{-1}$,][]{belli14lris, bezanson15}, suggesting that gas and stars have different distributions \citep{vandokkum15}.

\begin{figure*}[htbp]
\centering
\includegraphics[width=0.44\textwidth]{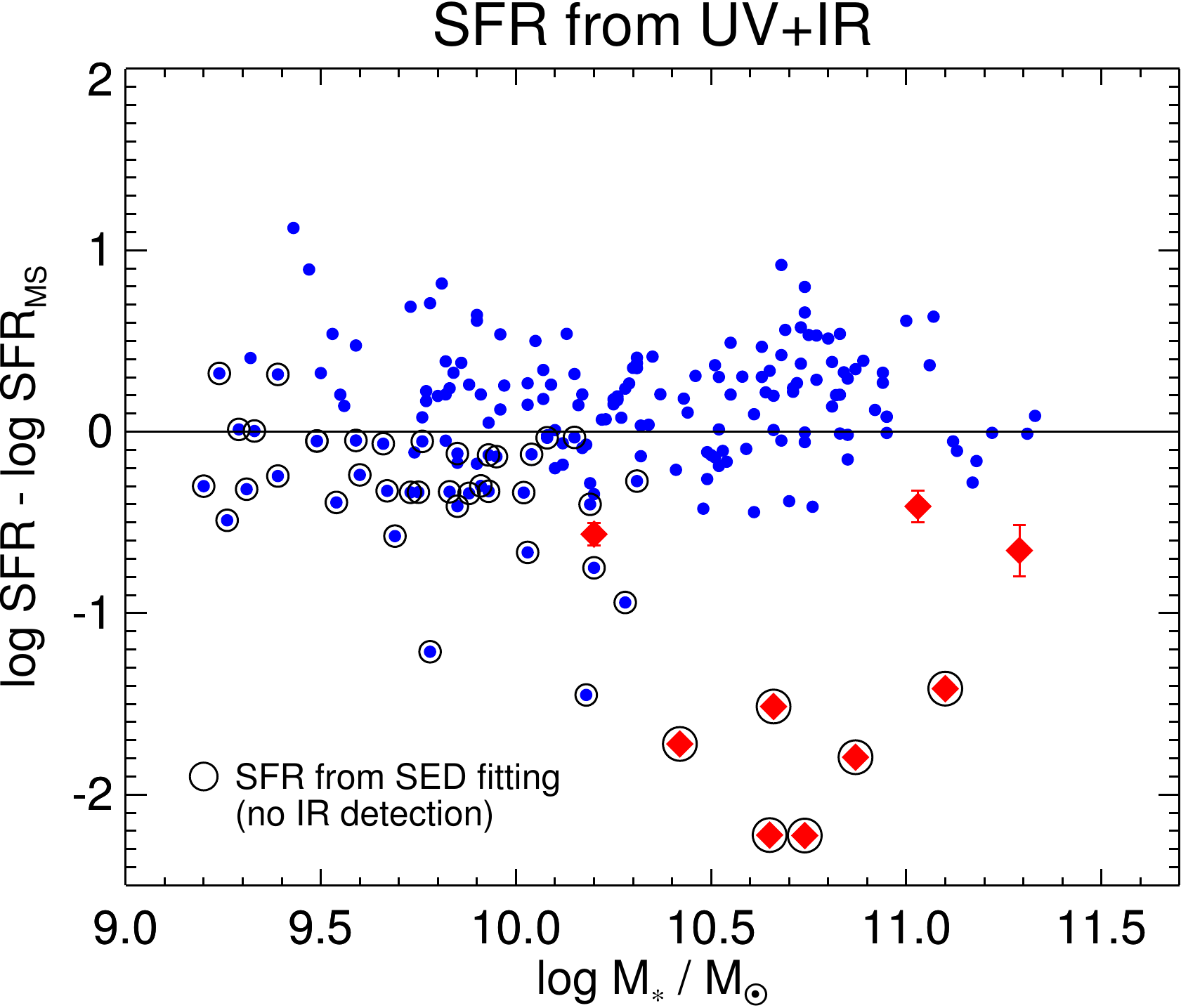}
\hspace{12mm}
\includegraphics[width=0.44\textwidth]{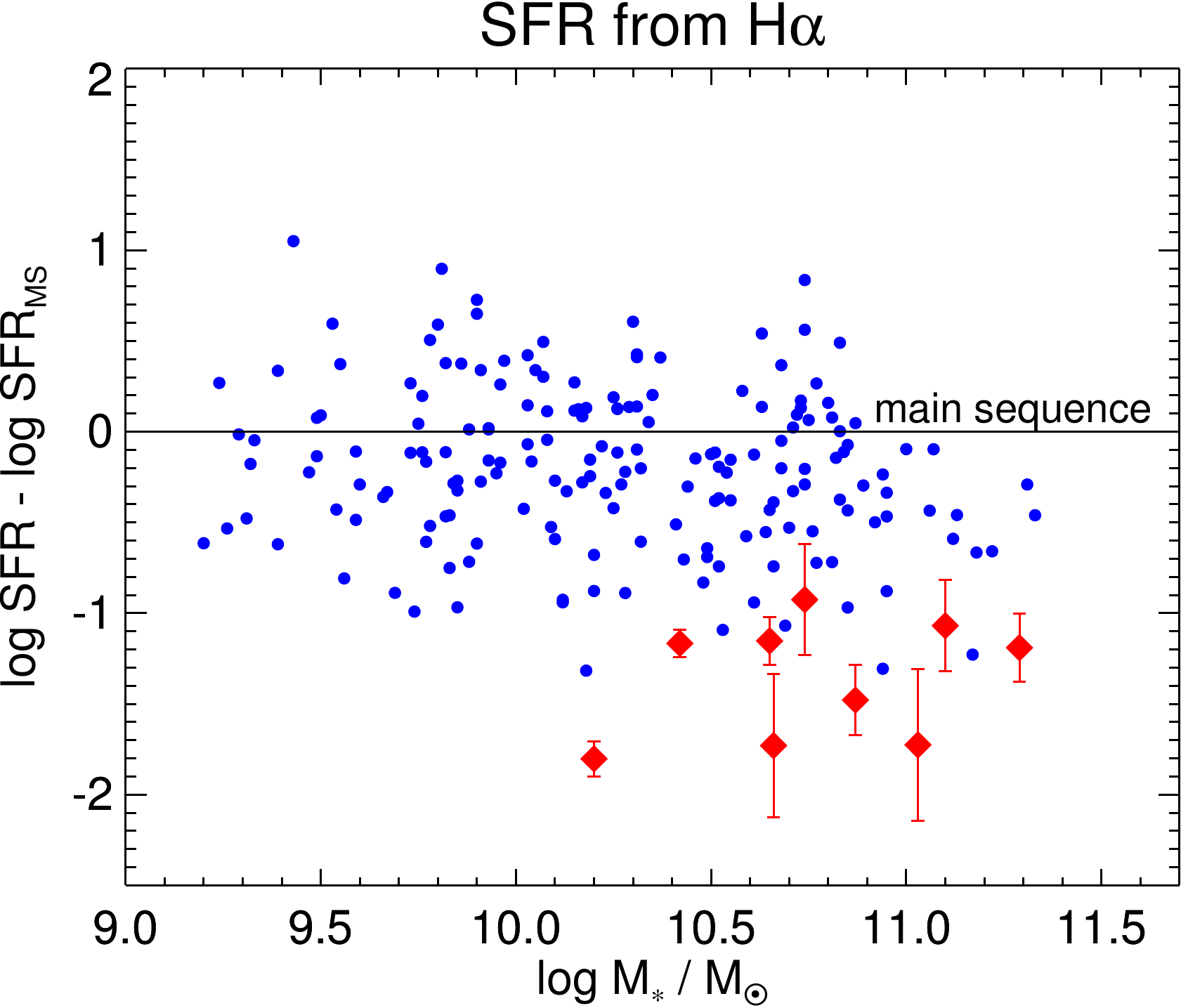}
\caption{Vertical distance from the main sequence \citep[as given by][]{whitaker14}, in logarithmic units, as a function of stellar mass. Star formation rates are derived from UV+IR luminosities (left panel) and dust-corrected \Halpha\ fluxes (right panel). The error bars in the right panel include conservative estimates of the uncertainty in the underlying \Halpha\ absorption and dust extinction (a factor of two for each contribution); in the left panel the SED-derived values are dominated by systematic uncertainties. Symbols and colors as in Figure \ref{fig:UVJ}; only galaxies at $z < 1.7$ and with \NII/\Halpha $ < 0.5$ are shown.}
\label{fig:mainseq}
\end{figure*}

\subsection{Gas Metallicity}

Measuring gas metallicity is challenging, especially when a limited set of strong emission lines are available. However, we can get a rough estimate of the \emph{relative} metallicity of the galaxies in our sample using the \NII/\Halpha\ line ratio, which rises monotonically with gas metallicity \citep[e.g.,][]{maiolino08}. As seen from the top and bottom panels of Figure \ref{fig:whan}, and from Figure \ref{fig:stacks}, the \NII-weak quiescent galaxies tend to have lower \NII/\Halpha\ ratios at fixed \Halpha\ equivalent width and fixed stellar mass compared to star-forming galaxies. The ionized gas in these quiescent galaxies is therefore more metal poor than what found in the majority of the star-forming objects of similar mass.

\subsection{Star Formation Rates}

We calculate star formation rates from the \Halpha\ fluxes using the \citet{kennicutt98} calibration converted to a \citet{chabrier03} IMF, and we obtain an average value of 1~\Msun/yr for the \NII-weak quiescent galaxies. However, part of the \Halpha\ emission can be hidden by dust. This should not be a large effect, since the SED fit shows that the attenuation for these objects is low, $A_V \lesssim 0.5$. We use the SED fit results to estimate the extra attenuation in HII regions following \citet{wuyts13}, and calculate dust-corrected star formation rates, that range from 0.2 to 7 \Msun/yr, and have a mean value of 1.5 \Msun/yr. We also estimate, from the best-fit SED model, the effect of underlying \Halpha\ stellar absorption. This is a negligible contribution in all but two object (U3-19342 and U4-22621), for which the absorbed flux is comparable to the detected emitted flux.

It is instructive to compare our results to the more commonly used method of deriving the star formation rate from the UV+IR luminosity. Figure \ref{fig:mainseq} shows the star formation rates, normalized to the value at the main sequence \citep{whitaker14}, as a function of stellar mass. In the left panel we show the star formation rates derived from the UV+IR luminosity, or from the SED fitting for those galaxies not detected at 24 $\mu$m (see \citealt{wuyts11} for details). Three of the nine quiescent galaxies are detected in the IR and their implied star formation rates are rather high, placing them near the main sequence. When using the more robust \Halpha-derived star formation rates, however, all quiescent galaxies lie one to two orders of magnitude below the main sequence, as shown in the right panel. We conclude that the UV+IR method can significantly overestimate the star formation rate of quiescent galaxies, confirming the result of photometric studies \citep[e.g.][]{utomo14, fumagalli14} and numerical simulations \citep[e.g.,][]{hayward14}, which suggest that old stellar populations can significantly contribute to the dust heating. Interestingly, infrared detections in our sample are anticorrelated with dust extinction and $U-V$ colors, confirming that the reddest quiescent galaxies are not hosting large amounts of dust-obscured star formation.

Another important result revealed by Figure \ref{fig:mainseq} is that the \Halpha-based main sequence is broader. Since \Halpha\ probes shorter timescales, this likely reflects a high degree of burstiness \citep[e.g.,][]{sparre17,faucher-giguere17}, but may also be due to variations in the dust extinction curve \citep{shivaei15}. As a result, the \Halpha-based main sequence forms a more continuous distribution with the quiescent population \citep[see, e.g.,][]{eales17}. \\


\vspace{-3mm}

\section{Discussion}
\label{sec:discussion}

We presented the discovery of nine massive quiescent galaxies at $z>0.7$ that host low-level star formation activity. The gas-phase metallicity of these objects is low, and many of them have one or more close companions, which in three cases are spectroscopically confirmed. We therefore conclude that these galaxies are fueled by recently accreted gas, either via inflows or gas-rich minor mergers.

Our observations thus suggest that these galaxies are caught during a rejuvenation event likely triggerred by interactions with gas-rich, lower-mass and lower-metallicity systems. This scenario is consistent with recent observations of a mild age gradient in the stellar populations of $z\sim1.5$ quiescent galaxies \citep{gobat17}. These systems can still be considered quiescent: assuming a constant star formation rate, we calculate an average increase of stellar mass due to star formation of only 10\% from $z\sim1$ to $z\sim0$, a value that is remarkably similar to the prediction of numerical simulations \citep{naab14}. This growth should be considered as an upper limit, because the gas accretion will probably decline with time. This is confirmed by the fact that massive ($\log \Mstar/\Msun > 11$) local ellipticals, which are the likely descendants of the [NII]-weak galaxies in our sample, show virtually no sign of recent star formation activity \citep{temi09,salim12}.

If these objects were already passively evolving before accreting gas, the rejuvenation event would need to be just strong enough to bring the broadband colors of the galaxies at the boundary of the UVJ diagram that separates the red sequence from the star-forming population.  Alternatively, if these objects had not yet been completely quiescent but were undergoing quenching, the transitional rest-frame colors could still be dominated by their intermediate-age stellar populations but could not explain the low inferred metallicities of the star-forming gas. Whether these galaxies were quiescent or still star-forming in their recent past can only be unveiled by a detailed analysis of their star formation histories. This requires a comprehensive study of the \Halpha\ emission together with the UV-to-IR photometry, and will be performed in a future work. In any case, it is worth noting that the rest-frame colors are remarkably effective at selecting truly quiescent galaxies that lack any evidence of emission lines.

Finally, we note that approximately half of these objects host large gas disks. This observation contributes to an emerging scenario in which rotation plays an important role in all stages of galaxy evolution at high redshift. Compact star-forming galaxies, which are thought to be close to quenching, host rotating ionized gas (\citealt{vandokkum15}; E. Wisnioski et al., in prep.), which is still present in at least some of the quiescent galaxies in our sample; finally, this rotation is partly maintained also in gas-poor quiescent galaxies, as revealed by studies of their stellar kinematics \citep{newman15, belli17rotation}. Our observations thus offer a unique view of the formation of massive quiescent systems.


\bibliography{sirio}

\end{document}